\documentstyle[preprint,prl,tighten,aps,epsfig]{revtex}    %

\begin{document}

\newcommand{\wt}{\widetilde}
\newcommand{\imag}{\Im {\rm m}}
\newcommand{\real}{\Re {\rm e}}
\newcommand{\tanb}{\tan \! \beta}
\newcommand{\cotb}{\cot \! \beta}
\newcommand{\mto}{m^2_{\tilde{t}_1}}
\newcommand{\mttt}{m^2_{\tilde{t}_2}}
\newcommand{\mbo}{m^2_{\tilde{b}_1}}
\newcommand{\mbt}{m^2_{\tilde{b}_2}}
\newcommand{\ghat}{\hat{g}^2}
\newcommand{\htop}{\left| h_t \right|^2}
\newcommand{\hb}{\left| h_b \right|^2}

\draft

\begin{flushright}
hep--ph/0106327
\end{flushright}

\begin{center}
\begin{Large}
{\bf Probing MSSM Higgs Sector CP Violation at a Photon Collider
\footnote{\it To be 
published in the proceedings of the `Theory Meeting on Physics at 
Linear Colliders', 15--17 March 2001, KEK, Japan.}
}\\ 
\vspace{0.9 cm}
\end{Large}

\begin{large}
Jae Sik Lee \\ \vspace{0.6 cm} 
\end{large}

Theory Group, KEK, Tsukuba, Ibaraki 305--0801, Japan \\ \vspace{0.6 cm}
\end{center}

\begin{center}
{\bf Abstract}
\end{center}

\begin{flushleft}
We study the phenomenological implications of the Higgs sector CP violation
at a photon collider. In our model, the CP violation
is radiatively induced by the non--trivial CP phases of the
third--generation scalar--quark sector in the MSSM.
We re--evaluate the $s$--channel resonance production cross sections 
and the polarization asymmetries of the neutral Higgs bosons 
based on the calculation of the mass matrix of the neutral Higgs bosons
which is valid for any values of the relevant SUSY parameters.
The CP properties of the Higgs bosons
can be precisely probed through their $s$--channel
resonance productions at a photon linear collider by exploiting
circularly and/or linearly polarized backscattered laser photons.
\end{flushleft}

\vskip 0.4cm




\section{Introduction}
\label{sec:introduction}

The soft CP--violating Yukawa
interactions in the minimal supersymmetric standard model (MSSM) cause
the CP--even and CP--odd neutral Higgs bosons to mix
via loop corrections \cite{AP,DEM,PW,CDL,CEPW}.
Although the mixing is a radiative effect, the induced CP violation
in the MSSM Higgs sector can be large enough to affect the
Higgs phenomenology significantly at present and future colliders
~\cite{AP,PW,CEPW,EXCP_FC,CL1,CL2,CL3,CLP,Bae,CL4,CDGL}.  \\

In the light of the possible large CP--violating mixing, we study the effects
of the CP--violating mixing on the $s$--channel
resonance production cross sections and 
the polarization asymmetries of the neutral Higgs bosons at a photon collider.
We studied the effects of the CP--violating mixing 
at a photon collider via their $s$--channel resonance productions
in Ref.~\cite{CLP} based on the mass matrix derived by Pilaftsis and
Wagner \cite{PW}. The mass matrix, however, is not applicable
for large squark mass splitting. 
In this work, we re--evaluate the production cross sections and the
polarization asymmetries with the mass matrix \cite{CDL} 
which is valid for any values of the soft--breaking parameters.
Our study clearly shows that the CP properties of Higgs bosons can be precisely
probed through their
$s$--channel resonance productions via photon--photon 
collisions by use of circularly and/or linearly polarized backscattered
laser photons at a TeV--scale linear $e^+e^-$ collider.  \\

This paper is organized as follows. In Sec.~\ref{sec:CP violation} we give 
a brief review of the calculation \cite{CDL}
of the loop--induced CP--violating mass matrix of
the three neutral Higgs bosons.  
In Sec. \ref{sec:photon},  we investigate in detail
the dependence of the total production rates and the three polarization 
asymmetries on the CP--violating phases.
Finally, we summarize our findings in Sec.~\ref{sec:conclude}.

\section{CP Violation in the MSSM Higgs Sector}
\label{sec:CP violation}

The loop--corrected mass matrix of the neutral Higgs bosons in the MSSM
can be calculated from the effective potential \cite{CW,OKADA}
\begin{eqnarray}
\label{e2}
V_{\rm Higgs}\hskip -0.3cm 
   &= \, \frac{1}{2}m_1^2\left(\phi_1^2+a_1^2\right)
     +\frac{1}{2}m_2^2\left(\phi_2^2+a_2^2\right)
     -\left|m_{12}^2\right|\left(\phi_1\phi_2-a_1 a_2\right) 
           \cos (\xi + \theta_{12}) \nonumber \\ 
   & +\left|m^2_{12}\right|
      \left(\phi_1 a_2 +\phi_2 a_1\right)\sin(\xi+\theta_{12})
     +\frac{\ghat}{8} {\cal D}^2 
     +\frac{1}{64\,\pi^2} {\rm Str} \left[
           {\cal M}^4 \left(\log\frac{{\cal M}^2}{Q^2} 
	                  - \frac{3}{2}\right)\right] \,,
\end{eqnarray}
with ${\cal D} = \phi_2^2 + a_2^2 - \phi_1^2 - a_1^2$,
$\ghat = (g^2+g'^2)/4$,
and $\phi_i$ and $a_i$ ($i=1,2$) are the real fields of the neutral
components of the two Higgs doublets:
\begin{eqnarray}
\label{e1}
H_1^0 = \frac{1} {\sqrt{2}} \left( \phi_1 + i a_1 \right)\,, \ \ \ \ \
H_2^0 = \frac {{\rm e}^{i \xi}} {\sqrt{2}} \left( \phi_2 + i a_2 \right)\,.
\end{eqnarray}
The parameters $g$ and $g'$ are the SU(2)$_L$ and U(1)$_Y$
gauge couplings, respectively, and  $Q$ denotes the
renormalization scale. All the tree--level parameters of the effective
potential (\ref{e2}) such as $m_1^2, \ m_2^2 $ and 
$m_{12}^2=\left|m^2_{12}\right|{\rm e}^{i\theta_{12}}$, are the
running parameters evaluated at the scale $Q$. The potential (\ref{e2}) is
then almost independent of $Q$ up to two--loop--order corrections. 
The super--trace is to be taken over all the bosons and fermions 
that couple to the Higgs fields. \\

The matrix ${\cal M}$ in Eq.~(\ref{e2}) is 
the field--dependent mass matrix of all modes that
couple to the Higgs bosons. The dominant contributions in the MSSM come from
third generation quarks and squarks because of their large Yukawa couplings.
The field--dependent masses of the third generation quarks are given by
\begin{eqnarray}
\label{e3}
m_b^2 = \frac{1}{2} |h_b|^2 \left( \phi_1^2 + a_1^2
\right)\,, \ \ \ \
m_t^2 = \frac{1}{2} |h_t|^2 \left( \phi_2^2 + a_2^2
\right), 
\end{eqnarray}
where $h_b$ and $h_t$ are the bottom and top Yukawa couplings, respectively. 
The corresponding squark mass matrices read:
\begin{eqnarray}
{\cal M}_{\tilde t}^2 &= \mbox{$ \left( \begin{array}{cc} 
m^2_{\wt Q} + m_t^2 - \frac{1}{8} \left( g^2 - \frac{g'^2}{3} \right) {\cal D}
\,\,\,\,\,\,\,\,\,\,
&
- h_t^* \left[ A_t^* \left(H_2^0 \right)^* + \mu H_1^0 \right] \\
- h_t \left[ A_t H^0_2 + \mu^* \left( H_1^0 \right)^* \right] &
m^2_{\wt U} + m_t^2 - \frac{g'^2}{6} {\cal D}
\end{array} \right)\,, $} \nonumber\\
\nonumber \\ 
{\cal M}_{\tilde b}^2 &= \mbox{$ \left( \begin{array}{cc} 
m^2_{\wt Q} + m_b^2 + \frac{1}{8} \left( g^2 + \frac{g'^2}{3} \right) {\cal D}
\,\,\,\,\,\,\,\,\,\,
&
- h_b^* \left[ A_b^*  \left( H_1^0 \right)^* + \mu H_2^0 \right] \\
- h_b \left[ A_b H_1^0 + \mu^* \left( H_2^0 \right)^* \right] &
m^2_{\wt D} + m_b^2 + \frac{g'^2}{12} {\cal D}
\end{array} \right)\,, $} 
\label{e4}
\end{eqnarray}
where $m^2_{\wt Q}, \ m^2_{\wt U}$ and
$m^2_{\wt D}$ are the real soft SUSY--breaking squark-mass
parameters, $A_b$ and $A_t$ are the
complex soft SUSY--breaking trilinear parameters, and $\mu$ is the complex
supersymmetric Higgsino mass parameter. \\

The mass matrix of the Higgs bosons (at vanishing external momenta)
is then given by the second derivatives of the potential
evaluated at its minimum point
\begin{eqnarray}
  \left(\phi_1,\, \phi_2,\, a_1,\, a_2\right)
=\left(\langle \phi_1\rangle,\, \langle \phi_2 \rangle,\,
       \langle a_1\rangle,\, \langle a_2 \rangle\right)
=\left(v\cos\beta,\, v\sin\beta,\, 0,\, 0\right),
\end{eqnarray}
where $v=(\sqrt{2}\, G_F)^{-1/2}\simeq 246 \ {\rm GeV}$. 
The massless state $G^0 = a_1 \cos \beta - a_2 \sin \beta$ is
the would--be--Goldstone mode to be absorbed by the $Z$ boson. 
We are thus left with a mass--squared matrix ${\cal M}_H^2$ for three 
physical states, 
$a \,(= a_1 \sin \beta + a_2 \cos \beta), \ \phi_1$ and $\phi_2$. 
The mass matrix is real and symmetric, i.e. it has 6 independent entries. 
The diagonal entry for the pseudoscalar component $a$ reads:
\begin{eqnarray}
\label{e10}
\left. {\cal M}^2_{H} \right|_{aa} = m_A^2 + \frac {3} {8 \pi^2}
\left\{ \frac { |h_t|^2 m_t^2 } { \sin^2 \beta} g(\mto,
\mttt) \Delta_{\tilde t}^2 +
\frac {|h_b|^2 m_b^2 } { \cos^2 \beta} g(\mbo,
\mbt) \Delta_{\tilde b}^2 \right\}\,,
\end{eqnarray}
where $m_A$ is the loop--corrected pseudoscalar mass in the CP invariant
theories.
The CP--violating entries of the mass matrix,
which mix $a$ with $\phi_1$ and $\phi_2$, are given by
\begin{eqnarray}
\label{e13}
\left. {\cal M}^2_H \right|_{a \phi_1}
   &=& \frac {3} {16 \pi^2} \left\{
       \frac { m_t^2 \Delta_{\tilde t} } {\sin \beta} \left[ g(\mto, \mttt)
       \left( X_t \cotb - 2 \htop R_t \right)
            - \ghat \cotb \log \frac{\mttt}{\mto} \right] \right. \\
   && \left. \hskip 1cm
     +\frac {m_b^2 \Delta_{\tilde b}} {\cos \beta} \left[ -g(\mbo,\mbt)
      \left( X_b + 2 \hb R_b' \right) + \left( \ghat - 2 \hb \right) \log
      \frac {\mbt} {\mbo} \right] \right\}, \nonumber\\
\left. {\cal M}^2_H \right|_{a \phi_2}
   &=& \frac {3} {16 \pi^2} \left\{
       \frac {m_t^2 \Delta_{\tilde t}} {\sin \beta} \left[ -g(\mto,\mttt)
       \left( X_t + 2 \htop R_t' \right)
            + \left(\ghat-2\htop\right)\log\frac{\mttt}{\mto} \right]\right.
\\
   && \left. \hskip 1cm
     +\frac { m_b^2 \Delta_{\tilde b} } {\cos \beta} \left[ g(\mbo, \mbt)
      \left( X_b \tanb - 2 \hb R_b \right)
            - \ghat\tanb\log \frac{\mbt}{\mbo} \right]\right\}\,, \nonumber
\end{eqnarray}
where $g(x,y)=2-[(x+y)/(x-y)]\log(x/y)$.
The size of these CP--violating entries is determined by
the re--phasing invariant quantities
\begin{eqnarray}
\label{e9}
\Delta_{\tilde t} = \frac { \imag(A_t \mu {\rm e}^{i \xi}) }
{\mttt - \mto} \,, \ \ \qquad
\Delta_{\tilde b} = \frac { \imag(A_b \mu {\rm e}^{i \xi}) }
{\mbt - \mbo} ,
\end{eqnarray}
which measure the amount of CP violation in the top and bottom squark--mass
matrices. In the CP--conserving limit, both  
$\Delta_{\tilde t}$ and $ \Delta_{\tilde b}$ vanish, leading to
$|m^2_{12}|\,\sin(\xi + \theta_{12})=0$.
The definition of the mass--squared $m^2_A$ and the dimensionless quantities
$X_{t,b}$, $R_{t,b}$ and $R^\prime_{t,b}$,
as well as the other CP--preserving entries of
the mass matrix squared ${\cal M}^2_{H}$,  can be found in Ref.~\cite{CDL}. 
The real and symmetric matrix ${\cal M}^2_H$ can now be
diagonalized with an orthogonal matrix $O$;
\begin{eqnarray}
\label{OMIX}
\left(\begin{array}{c}
       a      \\
       \phi_1 \\
       \phi_2
      \end{array}\right)\,=\,O\,\,
\left(\begin{array}{c}
       H_1 \\
       H_2 \\
       H_3
      \end{array}\right)\,.
\end{eqnarray}
Our convention for the three mass eigenvalues is
$m_{H_1}\leq m_{H_2}\leq m_{H_3}$.  \\

The loop--corrected neutral--Higgs--boson sector depends on
various parameters from the other sectors of the MSSM; 
$m_A$, which becomes the mass of the CP--odd Higgs boson if CP is
conserved, and $\tan\beta$ fix the tree--level Higgs potential; and
$\mu$, $A_t$, $A_b$ and the soft--breaking third generation sfermion
masses $m_{\tilde Q}$, $m_{\tilde U}$, and $m_{\tilde D}$, which fix
the third generation squark mass matrices. After minimization of the
potential the rephasing invariant sum $\theta_{12} + \xi$ of the
radiatively induced phase $\xi$ and the phase $\theta_{12}$ of the
soft breaking parameter $m^2_{12}$ is no longer an independent
parameter.
\footnote{As discussed in \cite{CDL}, $\xi$ and
$\theta_{12}$ are not separately physical parameters. For example, one
or the other can be set to zero in certain phase conventions for the
fields. Similar remarks hold for the phases of $A_t$, $A_b$ and
$\mu$. Altogether there are only three rephasing invariant
(i.e. physical) phases, which we write as $\theta_{12}+\xi$, $\arg(A_t
\mu {\rm e}^{i\xi})$ and $\arg(A_b \mu {\rm e}^{i\xi})$. The
minimization of the potential fixes one of these combinations, leaving
two independent physical phases as free input parameters.} 
The physically meaningful CP phases in the Higgs sector are thus the
phases of the re--phasing invariant combinations $A_t\, \mu \, {\rm
e}^{i\xi}$ and $A_b\, \mu \, {\rm e}^{i\xi}$ appearing in
Eq.~(\ref{e9}). The neutral--Higgs--boson mixing also depends on
the complex gluino--mass parameter $M_{\wt g}$ through one--loop corrections 
to the top and bottom quark masses \cite{SUSYHBB}.  \\

Noting that the size of the radiative Higgs sector CP violation is determined
by the rephasing invariant combinations $A_t \mu {\rm e}^{i\xi}$ and 
$A_b \mu {\rm e}^{i\xi}$, see Eq. (\ref{e9}), 
we take for our numerical analysis the following set of parameters:
\begin{eqnarray}
&& |A_t|=|A_b| = 1~{\rm TeV}\,, 
\hspace{1.9 cm} |\mu|=2~{\rm TeV}\,,\nonumber \\
&& m_{{\widetilde Q},{\widetilde U},{\widetilde D}}=|M_{\widetilde g}|
= 0.5~{\rm TeV}\,,
\qquad {\rm Arg}(M_{\widetilde g})=0\,,
\label{eq:PARA}
\end{eqnarray}
under the constraint;
\begin{equation}
\Phi \equiv {\rm Arg}(A_t\mu {\rm e}^{i\xi}) = 
{\rm Arg}(A_b\mu {\rm e}^{i\xi})\,.
\end{equation}
We vary the common phase $\Phi$ as well as $m_A$ and $\tan\beta$ in 
the following numerical studies. Our choice of relatively large magnitudes of
$|A_t\mu|=|A_b\mu|$ enhances the effects of the CP violation 
in the MSSM Higgs sector.  
\\

The CP--violating phase could weaken the LEP lower 
limit on the lightest Higgs boson mass significantly \cite{KW,CEPW2}. 
In our analysis we show our results when the lightest Higgs--boson mass is
above 70 GeV. 

\section{Photon Linear Collider}
\label{sec:photon}

One of the cleanest determinations of the neutral Higgs sector CP violation 
in the MSSM can be achieved by observing the CP properties of all three 
neutral Higgs particles directly. In this light, the $s$--channel resonance 
production of neutral Higgs bosons in $\gamma\gamma$ collisions 
\cite{TWOPHOTON} has long been recognized as an important instrument to study 
the CP properties of Higgs particles \cite{GG,LINEAR} at a linear $e^+e^-$ 
collider (LC) by use of polarized high energy laser lights obtained by 
Compton back--scattering of polarized laser light off the electron and
positron beams \cite{GKPST}. In this section, we demonstrate that
the polarized back--scattered laser photons at a TeV--scale LC enable us to 
investigate the CP violation of the Higgs sector in the MSSM through 
$s$--channel Higgs--boson production via $\gamma\gamma$ collisions 
in detail including its dependence on the relevant SUSY parameters \cite{CLP}.
\\

In the presence of the CP--violating neutral Higgs--boson mixing, 
the amplitude for the two--photon fusion process 
$\gamma\gamma\rightarrow H_i$ ($i=1,2,3$) can be written 
in terms of two (complex) form factors $S^\gamma_i(s)$ and $P^\gamma_i(s)$ as
\begin{eqnarray} \label{me}
{\cal M}(\gamma\gamma\rightarrow H_i)=
\sqrt{s}\frac{\alpha}{4\pi}\bigg\{S^\gamma_i(s)
\left(\epsilon_1\cdot\epsilon_2-\frac{2}{s}k_1\cdot\epsilon_2
\,k_2\cdot\epsilon_1\right)
 -P^\gamma_i(s)\frac{2}{s}
\epsilon_{\mu\nu\rho\sigma}\,\epsilon_1^\mu \epsilon_2^\nu k_1^\rho
k_2^\sigma\bigg\}\,, 
\end{eqnarray}
where $s$ is the c.m. energy squared of two colliding photons.
In the two--photon c.m. coordinate system with one photon momentum $\vec{k}_1$
along the positive $z$ direction and the other one $\vec{k}_2$ along the 
negative $z$ direction, the wave vectors $\epsilon_{1,2}$ of two photons 
are given by
\begin{eqnarray} \label{wv}
\epsilon_1(\lambda)=\epsilon^*_2(\lambda)
                   =\frac{1}{\sqrt{2}}\left(0,-\lambda,-i,0\right)\,.
\end{eqnarray}
where $\lambda=\pm 1$ denote the right and left photon helicities, respectively.
In the MSSM with radiative CP--violating Higgs mixing, the scalar and 
pseudoscalar form factors are given by
\begin{eqnarray}
S^\gamma_i(s)&=&2N_C\sum_{f=t,b}e_f^2\left\{
g_{sf}^i\frac{\sqrt{s}}{m_f} F_{sf}(\tau_{sf})
+\frac{1}{4}\sum_{j=1,2}
g_{\tilde{f}_j\tilde{f}_j}^i
\frac{\sqrt{s}}{m_{\tilde{f}_j}^2} F_0(\tau_{s\tilde{f}_j})\right\}
 \nonumber \\
&&\hspace{0.5 cm}+
\frac{g \sqrt{s}}{2m_W}(c_\beta O_{2,i}+s_\beta O_{3,i})F_1(\tau_{sW})+
\frac{v \sqrt{s} C_i}{2 M_{H^\pm}^2} F_0(\tau_{sH^\pm})
\,, \nonumber \\
P^\gamma_i(s)&=&2N_C\sum_{f=t,b} e_f^2g_{pf}^i\frac{\sqrt{s}}{m_f}
F_{pf}(\tau_{sf})
 \,,
\end{eqnarray}
with $\tau_{sx}=s/4m_x^2$ and $N_C=3$.  The definitions of 
the four form factors $F_{0}$, $F_{sf}$, $F_{pf}$, and $F_1$ and
the couplings of the neutral Higgs
bosons to fermions, sfermions, and the charged Higgs--boson and $W$--boson
pairs can be found in Ref.~\cite{CLP}.
The amplitude for the production of the lightest Higgs boson, mass of which is
bounded below about 130 GeV in the MSSM, is dominated by the contribution 
from $W$--boson loop through the scalar form factor $S^\gamma_1(M_{H_1}^2)$. 
\\

Inserting the wave vectors (\ref{wv})
into Eq.~(\ref{me}) we obtain the production
helicity amplitude for the photon fusion process as follows 
\begin{eqnarray}
{\cal M}_{\lambda_1\lambda_2}=-\sqrt{s}\frac{\alpha}{4\pi}
     \left\{ S^\gamma_i(s)\,\delta_{\lambda_1\lambda_2}
          +i \lambda_1 P^\gamma_i(s)\delta_{\lambda_1\lambda_2}\right\}\,, 
\label{hamp}
\end{eqnarray}
with $\lambda_{1,2}=\pm$. For the $s$--channel resonance production of the
neutral Higgs boson $H_i$, the c.m. energy squared $s$ is to be replaced with 
$m_{H_i}^2$. And the absolute polarized amplitude squared is given by
\begin{eqnarray}
\overline{\left|{\cal M}\right|^2}=
\overline{\left|{\cal M}\right|_0^2}\Bigg\{
\left[1+\zeta_2\tilde{\zeta}_2\right]
+{\cal A}_1\left[\zeta_2+\tilde{\zeta}_2\right]
+{\cal A}_2\left[\zeta_1\tilde{\zeta}_3+\zeta_3\tilde{\zeta}_1\right]
-{\cal A}_3\left[\zeta_1\tilde{\zeta}_1-\zeta_3\tilde{\zeta}_3\right] \Bigg\}
\,,
\end{eqnarray}
where $\{\zeta_i\}$ are the Stokes parameters describing 
the polarization transfer from the laser light to the high energy photons; 
$\zeta_2$ is the
degree of circular polarization and $\{\zeta_3,\zeta_1\}$ the degree of
linear polarization transverse and normal to the plane defined by the
electron direction and the direction of the maximal linear polarization
of the initial laser light.  
To acquire the high sensitivity to CP violation, it is necessary to control
both the energy of the initial laser light and its
degrees of the circular and transverse polarization \cite{GKPST,CLP}.
The unpolarized amplitude squared
$\overline{\left|{\cal M}\right|_0^2}$ is given by
\begin{equation}
\overline{\left|{\cal M}\right|_0^2}\,(\gamma\gamma\rightarrow H_i)=
\frac{M_{H_i}^2}{2}\left(\frac{\alpha}{4\pi}\right)^2
\left\{|S^\gamma_i(M_{H_i}^2)|^2+|P^\gamma_i(M_{H_i}^2)|^2\right\}\,,
\end{equation}
and three polarization asymmetries ${\cal A}_i$ ($i=1,2,3$) are defined in
terms of the helicity amplitudes and expressed in terms of the form factors
$S^\gamma_i$ and $P^\gamma_i$ as
\begin{eqnarray}
{\cal A}_1&=&\frac{\left|{\cal M}_{++}\right|^2-\left|{\cal M}_{--}\right|^2}
        {\left|{\cal M}_{++}\right|^2+\left|{\cal M}_{--}\right|^2}=
\frac{2{\cal I}\left[S^\gamma_i(M_{H_i}^2)P^{\gamma *}_i(M_{H_i}^2)\right]}
{|S^\gamma_i(M_{H_i}^2)|^2+|P^\gamma_i(M_{H_i}^2)|^2}\,,
\nonumber \\ \nonumber \\
{\cal A}_2&=&\frac{2{\cal I}({\cal M}_{--}^*{\cal M}_{++})}
        {\left|{\cal M}_{++}\right|^2+\left|{\cal M}_{--}\right|^2}=
\frac{2{\cal R}\left[S^\gamma_i(M_{H_i}^2)P^{\gamma *}_i(M_{H_i}^2)\right]}
{|S^\gamma_i(M_{H_i}^2)|^2+|P^\gamma_i(M_{H_i}^2)|^2}\,,
\nonumber \\ \nonumber \\
{\cal A}_3&=&\frac{2{\cal R}({\cal M}_{--}^*{\cal M}_{++})}
        {\left|{\cal M}_{++}\right|^2+\left|{\cal M}_{--}\right|^2}=
\frac{|S^\gamma_i(M_{H_i}^2)|^2-|P^\gamma_i(M_{H_i}^2)|^2}
{|S^\gamma_i(M_{H_i}^2)|^2+|P^\gamma_i(M_{H_i}^2)|^2}\,.
\end{eqnarray}
In the CP--invariant theories, the two form factors $S^\gamma_i$ and
$P^\gamma_i$ can not coexist. In other words,
non--zero ${\cal A}_{1,2}$ and/or $\left|{\cal A}_3\right|<1$ indicate the
CP violation. \\

\begin{figure*}
\begin{center}
\hbox to\textwidth{\hss\epsfig{file=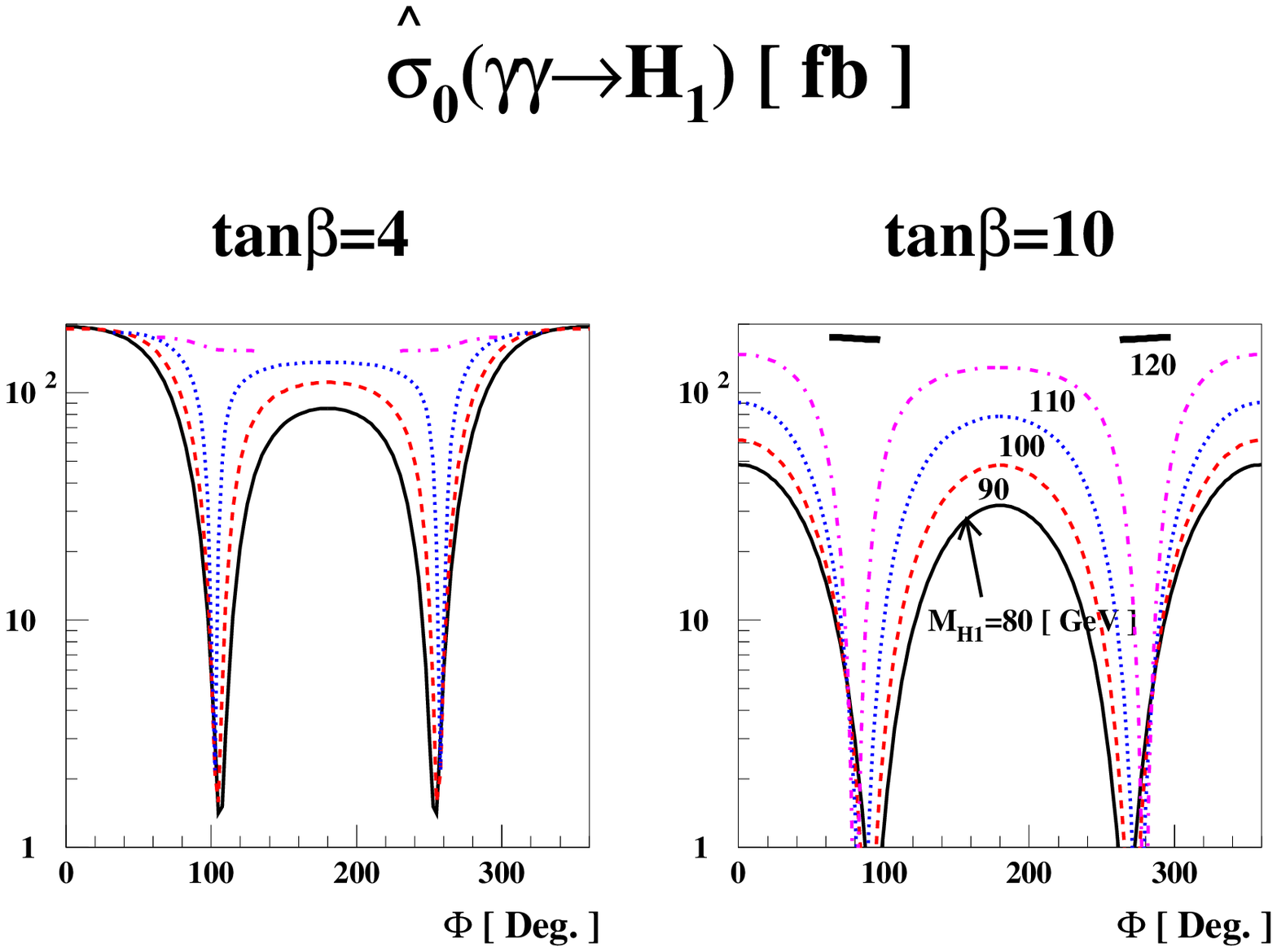,width=18cm,height=18cm}\hss}
\end{center}
\caption{
The unpolarized cross section of the lightest Higgs boson
in units of fb 
as a function of $\Phi$ for four ($\tan\beta=4$) and five
($\tan\beta=10$) values of $m_{H_1}$: $m_{H_1}=80$ GeV (solid line),
90 GeV (dashed line), 100 GeV (dotted line), 110 GeV (dash--dotted line), and
120 GeV (thick solid line). 
We take the parameter set Eq.~(\ref{eq:PARA}) with
$\tan\beta=4$ (left) and $\tan\beta=10$ (right).
}
\label{pph1}
\end{figure*}

The unpolarized cross section for the $s$--channel Higgs--boson production is
given by
\begin{equation}
\hspace{-1 cm}
\sigma(\gamma\gamma\rightarrow H_i)=\frac{\pi}{M_{H_i}^4}
\overline{\left|{\cal M}\right|_0^2}\,\delta\left(1-\frac{M_{H_i}^2}{s}\right)
\equiv {\hat \sigma}_0(\gamma\gamma\rightarrow H_i)
\delta\left(1-\frac{M_{H_i}^2}{s}\right)\,.
\end{equation}
Figure~\ref{pph1} shows the unpolarized cross section
of the lightest Higgs boson, 
$\hat{\sigma}_0(\gamma\gamma\rightarrow H_{1})$,
in units of fb 
as a function of $\Phi$ for four ($\tan\beta=4$) and five
($\tan\beta=10$) values of $m_{H_1}$: $m_{H_1}=80$ GeV (solid line),
90 GeV (dashed line), 100 GeV (dotted line), 110 GeV (dash--dotted line), and
120 GeV (thick solid line). 
We take the parameter set Eq.~(\ref{eq:PARA}) with
$\tan\beta=4$ (left) and $\tan\beta=10$ (right).
The unpolarized cross section for the lightest
Higgs boson strongly depends on the CP phase $\Phi$ as well as $m_{H_1}$.
The cross section is larger for larger values of $m_{H_1}$.
Note that this cross section is highly suppressed around $\Phi=100^{\rm o}
\,(90^{\rm o})$ and $260^{\rm o}\,(270^{\rm o})$ for $\tan\beta=4\,(10)$.
This is because the lightest Higgs boson contains a large admixture of 
CP--odd state $a$ in this region
and the $H_1W^\pm W^\mp$ coupling is significantly suppressed. \\

\begin{figure*}
\begin{center}
\hbox to\textwidth{\hss\epsfig{file=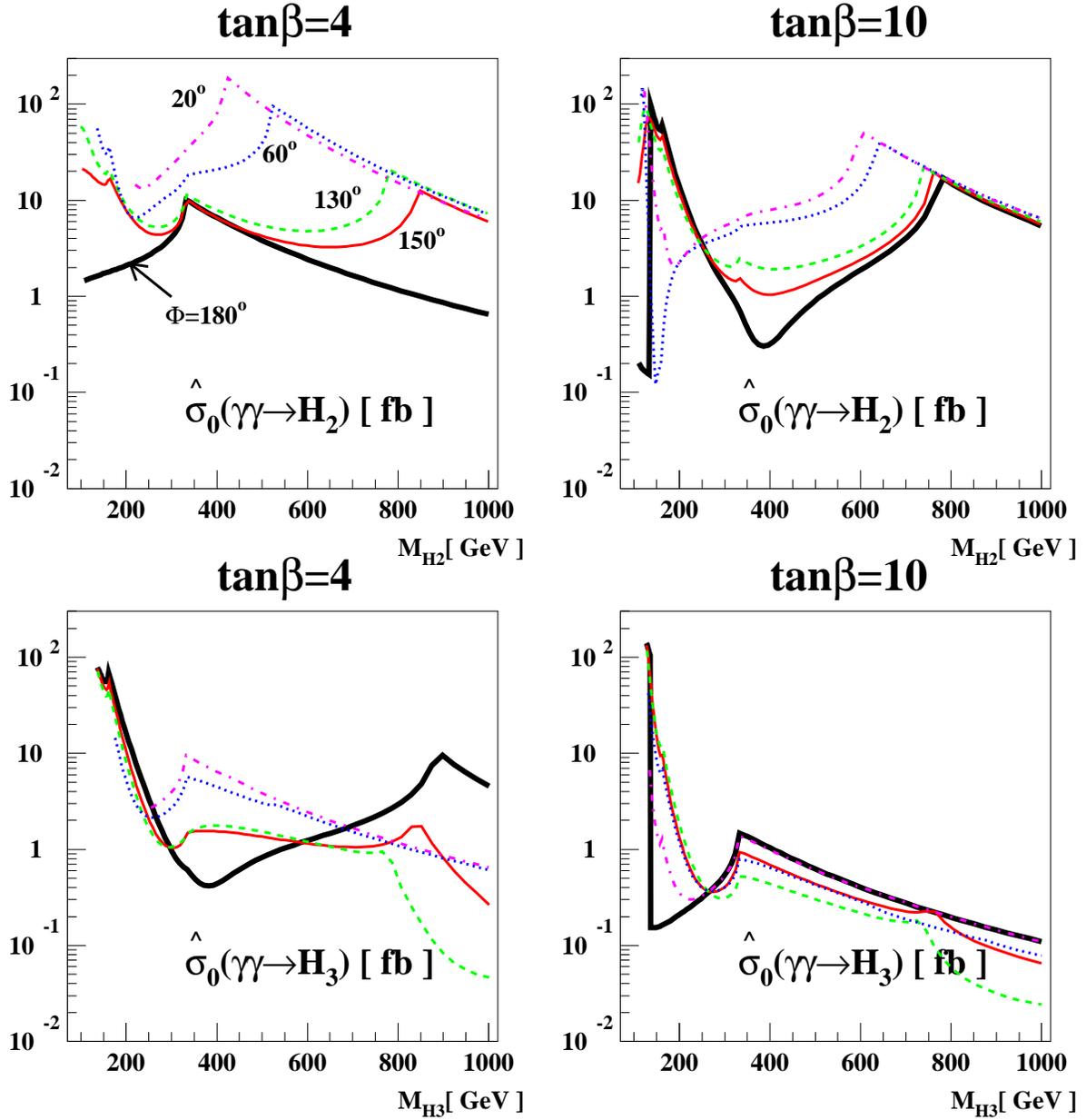,width=18cm,height=18cm}\hss}
\end{center}
\caption{
The unpolarized cross sections of the $H_2$ (upper) and $H_3$ (lower)
in units of fb as a function of each Higgs--boson mass for five values of
$\Phi$: $\Phi=180^{\rm o}$ (thick solid line), 150$^{\rm o}$ (solid line),
130$^{\rm o}$ (dashed line), 60$^{\rm o}$ (dotted line), and
20$^{\rm o}$ (dash--dotted line).
We take the parameter set Eq.~(\ref{eq:PARA}) with
$\tan\beta=4$ (left) and $\tan\beta=10$ (right).
}
\label{cs23}
\end{figure*}

Figure~\ref{cs23} shows the unpolarized cross sections
of the heavier two Higgs bosons,
$\hat{\sigma}_0(\gamma\gamma\rightarrow H_{2,3})$,
in units of fb as a function of each Higgs--boson mass for five values of
$\Phi$: $\Phi=180^{\rm o}$ (thick solid line), 150$^{\rm o}$ (solid line),
130$^{\rm o}$ (dashed line), 60$^{\rm o}$ (dotted line), and
20$^{\rm o}$ (dash--dotted line).
We take the parameter set Eq.~(\ref{eq:PARA}) with
$\tan\beta=4$ (left) and $\tan\beta=10$ (right).
The upper two frames are for the intermediate Higgs boson and 
the lower ones for the heaviest Higgs boson.
One can observe that these cross sections strongly depends on the CP 
phase $\Phi$. The behavior of the cross sections
can be understood by taking into 
account the $\Phi$ dependence of the couplings of the corresponding
Higgs boson to the fermion bilinear,
the diagonal sfermion pair, and
the $W$--boson and the charged Higgs--boson pairs. 
For example, let's closely look into the left--upper frame of Fig.~\ref{cs23} 
with $\tan\beta=4$ for the production of the second lightest Higgs 
boson $H_2$. For $\Phi=180^{\rm o}$, $H_2$ is CP--odd. In this case 
the couplings of $H_2$ to the diagonal sfermion pairs, 
the charged Higgs--boson and $W$--boson pairs vanish and
the production cross section
gets contributions only from the fermionic loops dominated by 
the top quark for $\tan\beta=4$. 
The mass dependence of the cross section comes from the form
factor $F_{pf}(m_{H_2}^2/4\,m_t^2)$ 
peaked at $m_{H_2}=2\,m_t$. When the CP phase $\Phi$ differs from
$180^{\rm o}$, $H_2$ becomes to contain the CP--even states. In other
words, the cross section starts to get contributions from $W$--boson and
sfermion loops. The $W$--boson--loop contribution is peaked at 
$m_{H_2}=2\,m_W$ and
the sfermion contributions, which is dominated by the lightest top squark
$\tilde{t}_1$, peaked at $m_{H_2}=2\,m_{\tilde{t}_1}$.
Note that the mass of the lightest top squark also depends on $\Phi$. The
lightest top--squark mass becomes lighter when $\Phi$ decrease from
$180^{\rm o}$.
Usually, the contribution from the charged Higgs--boson loops is suppressed
compared to the other three kinds of contributions due to
$m_{H^{\pm}}\sim m_{H_{2,3}}$.
\\

\begin{figure*}
\begin{center}
\hbox to\textwidth{\hss\epsfig{file=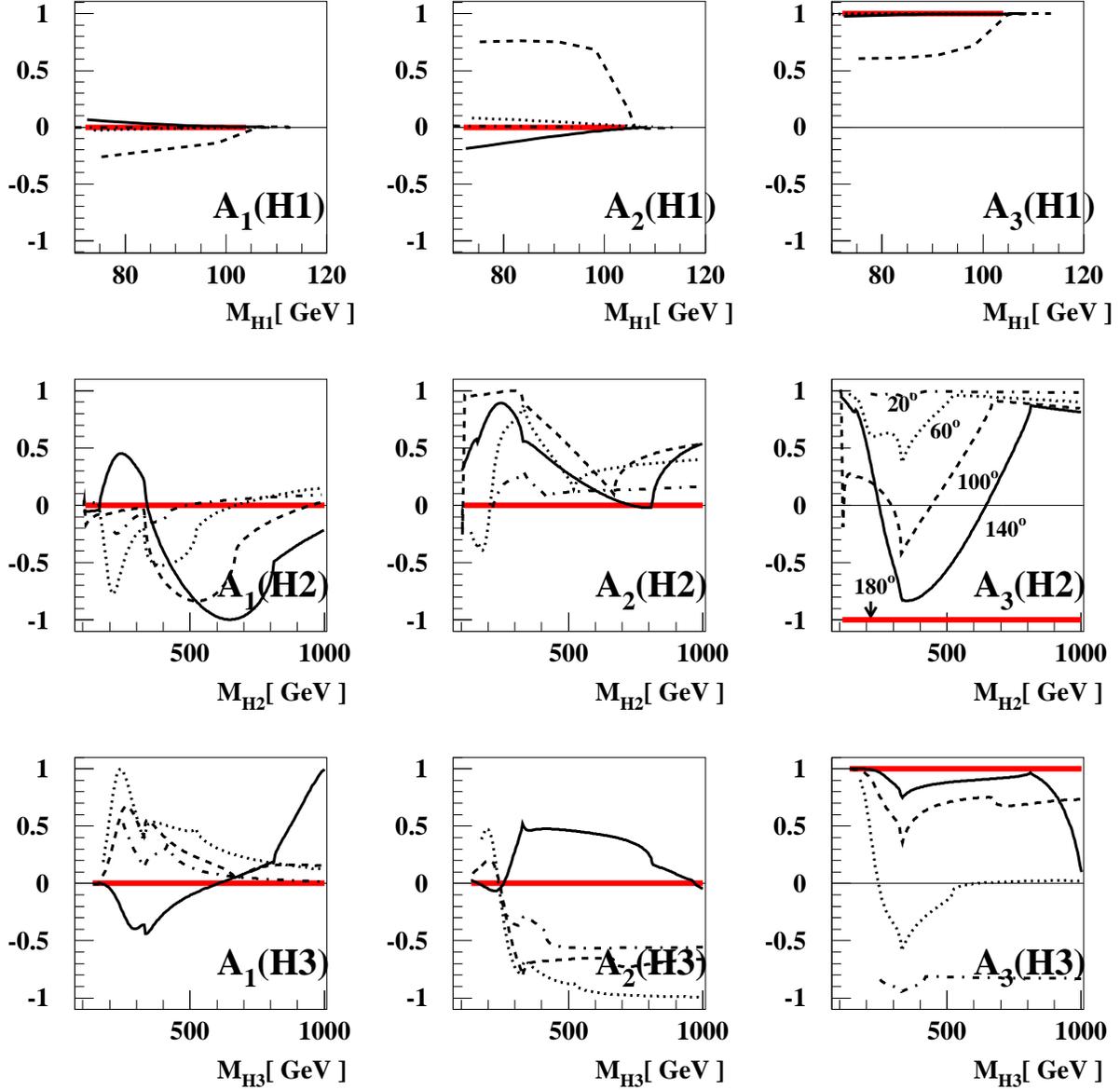,width=18cm,height=18cm}\hss}
\end{center}
\caption{The polarizatin asymmetries ${\cal A}_1$(left column),
${\cal A}_2$(middle column), and ${\cal A}_3$(right column) as functions of each
Higgs--boson mass for $\tan\beta=4$ for five values of $\Phi$;
$\Phi=180^{\rm o}$(thick solid line), 
$\Phi=140^{\rm o}$(solid line), $\Phi=100^{\rm o}$(dashed line), 
$\Phi=60^{\rm o}$(dotted line), and $\Phi=20^{\rm o}$(dash--dotted line). 
The upper 3 frames are for 
$H_1$ and the middle 3 ones for $H_2$, and the lower 3 ones for $H_3$.
}
\label{asym4}
\end{figure*}

\begin{figure*}
\begin{center}
\hbox to\textwidth{\hss\epsfig{file=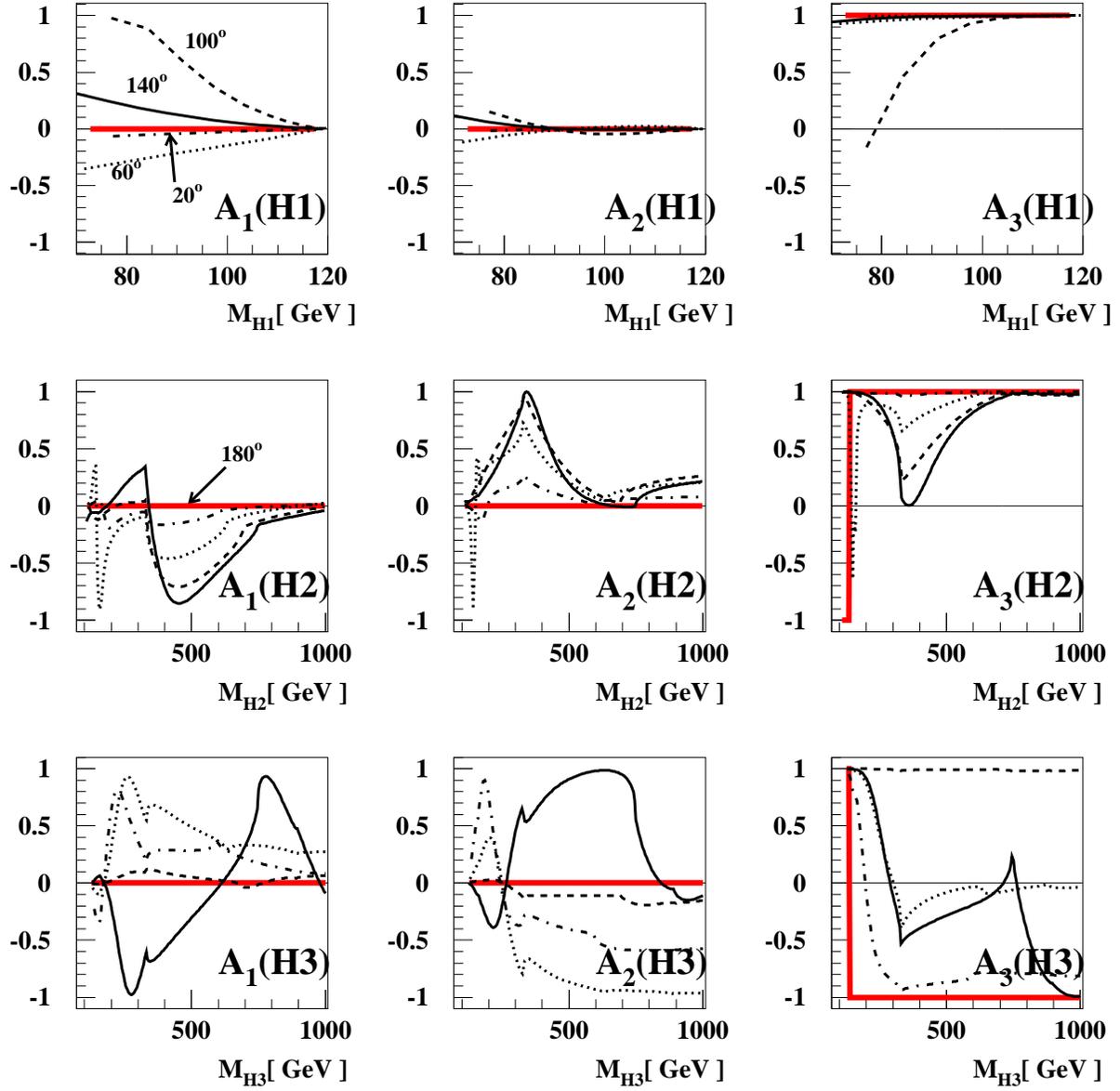,width=18cm,height=18cm}\hss}
\end{center}
\caption{The same as Figure \ref{asym4} but with $\tan\beta=10$. }
\label{asym10}
\end{figure*}

Figures~\ref{asym4} and \ref{asym10} shows three polarization asymmetries as
functions of each Higgs--boson mass for $\tan\beta=4$ and $\tan\beta=10$,
respectively. 
Noting that these polarization asymmetries satisfy the relations
\begin{equation}
{\cal A}_{1,2}(\Phi)=-{\cal A}_{1,2}(360^{\rm o}-\Phi)\,, \hspace{0.3 cm}
{\cal A}_{3}(\Phi)=+{\cal A}_{3}(360^{\rm o}-\Phi)\,,
\end{equation}
we choose five values of $\Phi$ less than $180^{\rm o}$
to show the dependence of the asymmetries on the CP--violating phase:
$\Phi=180^{\rm o}$ (thick solid line), 140$^{\rm o}$ (solid line),
100$^{\rm o}$ (dashed line), 60$^{\rm o}$ (dotted line), and
20$^{\rm o}$ (dash--dotted line). 
In the CP--conserving limit ($\Phi=180^{\rm o}$),
the polarization asymmetries ${\cal A}_{1,2}$ vanish and
the asymmetry ${\cal A}_3$ takes one of the values +1 (CP--even) or 
-1 (CP--odd)  depending on the CP--parity of the Higgs bosons in this limit.
For $\tan\beta=10$, the polarization asymmetry ${\cal A}_1$ of the lightest
Higgs boson is quite sensitive to the CP--phase $\Phi$. 
The other polarization asymmetries of the lightest Higgs boson are not much
different from those in the CP--conserving limit except $\Phi=100^{\rm o}$.
But, we note that all of three polarization asymmetries 
of the two heavier Higgs bosons significantly
differ from those of the CP--conserving theory even for small CP violation
with $\Phi=20^{\rm o}$ and $140^{\rm o}$ independently of $\tan\beta$.

\section{Conclusions}
\label{sec:conclude}
Based on the calculation of the mass matrix of the neutral Higgs bosons
which is valid for any values of the relevant SUSY parameters, we have
re--evaluated the $s$--channel resonance production cross sections 
and the polarization asymmetries of the neutral MSSM Higgs bosons
in the presence of the non--trivial CP--violating mixing among them.
The cross section of the lightest Higgs boson which is dominated by the
$W$--boson loop can be highly suppressed when the lightest Higgs boson is
almost CP--odd. For the heavier Higgs bosons, the cross sections strongly
depends on the CP--violating mixing.
The polarization asymmetries of two heavier Higgs bosons are very sensitive to
the non--trivial CP phases.
Our detailed analysis has clearly shown that collisions 
of polarized photons can provide a significant opportunity for detecting 
CP violation in the MSSM Higgs sector induced at the
loop level.

\section*{Acknowledgments}

I wish to thank my collaborators, Eri Asakawa, Seong--Youl Choi, Manuel
Drees, Benedikt Gaissmaier, and Kaoru Hagiwara.
This work was supported
by the Japan Society for the Promotion of Science (JSPS).

\vskip 1.5cm


\vfil\eject

\end{document}